# Cancer multi-omics joint dimension reduction leads to reliable and interpretable survival predictions

Vincent Le Goff[1,*], Vincent Guillemot[2], Cathy Philippe[3], Wiam Ahdidou[1], Gwendoline Mendes[1], Jean-François Deleuze[1,4], Edith Le Floch[1] & Arnaud Gloaguen[1]

[1] *Centre National de Recherche en Génomique Humaine (CNRGH), CEA, Université Paris-Saclay, Evry-Courcouronnes*

[2] *Institut Pasteur, Université Paris Cité, Bioinformatics and Biostatistics Hub, Paris*

[3] *Neurospin, CEA, Université Paris-Saclay, Gif-sur-Yvette*

[4] *Centre d'Etude du Polymorphisme Humain (CEPH) Fondation Jean Dausset, Paris, France*

[*] *vincent.legoff@cnrgh.fr*

**Abstract** Multi-omics studies are becoming more and more frequent, and are especially suited to decrypt the complexity of dysregulation happening in cancer. In this context, improving prognostication through interpretable methods could provide better patient stratification and help identify new prognostic biomarkers and therapeutic targets. Multi-omics data are however very often high-dimensional (few patients/high number of variables). A possible solution to high-dimensionality is Joint Dimension Reduction (JDR) methods, which compute a low-dimensional latent space from multi-omics data. In this work, we benchmark 12 JDR methods on 18 different TCGA cancer datasets. Survival prediction is achieved through a two-step process: the estimation of a low-dimensional multi-omics latent space followed by survival prediction with or without additional clinical data. We had to equip 4 methods with the ability to predict the survival of new patients. Results reveal that JDR methods can improve survival prediction with clinical data, and match the predictive performance of clinical data alone, with a high variability between different cancers. Among JDR methods, JIVE and RGCCA show superior performance. In a subsequent bootstrap analysis, JDR methods identify reliable prognostic biomarkers across cancers. We propose a clinically relevant interpretation of such markers on Low-Grade Glioma. Reproducible codes are available on `https://github.com/vlegoff/MultiOBenchmark`.

## 1 Introduction

The lowering cost of sequencing has facilitated the acquisition of multiple omics modalities on a single cohort, giving rise to multi-omics datasets. This type of datasets could improve the understanding of diseases by allowing the identification of processes dysregulated in multiple omics. In the context of cancer specifically, The Cancer Genome Atlas (TCGA) (`https://www.cancer.gov/tcga`) has produced multi-omics data with genomic (somatic mutations, copy number variants), transcriptomic (gene expression, miRNA expression), epigenomic (DNA methylation) and proteomic information for thousands of samples across multiple cancers. As such, TCGA constitutes a valuable resource to further the understanding of cancer. The main outcome of interest of many cancer studies is often overall survival, or progression-free survival. Resources such as TCGA and other multi-omics cohorts could help improve the accuracy of such predictions, which can help build better molecular-based stratifications into disease subtypes and discover new prognostic biomarkers and therapeutic targets.

However, analyzing multi-omics data is a challenging task, first and foremost because of their inherent high-dimensionality. Multi-omics datasets indeed often contain upward of 10 000 or 100 000 predictors, greatly outnumbering the sample size. Multiple methods have been proposed to tackle this challenge [1], ranging from classical machine learning to deep learning methods. Classical machine learning includes the Cox model [2] and its regularized extensions: IPF-LASSO [3], Priority-LASSO [4], boosting methods based on the Cox model including CoxBoost [5] and random forest based methods such as BlockForest [6]. These methods have been benchmarked by Herrmann et al. [7], who have used 18 TCGA datasets to show that when clinical variables are included alongside multi-omics data no method performs significantly better on average than the baseline model on clinical data only. Of notice, BlockForest appears to be among the best performing methods once they aggregate results across the 18 cancer datasets. Deep Learning (DL) methods have also been developed for single-omic survival analysis, such as DeepSurv [8] and Cox-nnet [9]. Their adaptation to multi-omics analysis have been benchmarked by Wissel et al. [10] in the context of cancer prognostication and have shown poor resistance to noise, while providing performances similar to non-DL methods, such as BlockForest. DL architectures have also been developed specifically to integrate multi-omics data, most notably CustOmics [11]. In the very same context, DL models similar to CustOmics lead to the best prognosis precisions among DL methods, while still similar to BlockForest's performances [12].

Although none of the methods cited above reliably outperforms models based on clinical data only, Herrmann et al. [7] show that methods that leverage the modality structure of multi-omics data outperform classical ones. However, this study leaves out a class of commonly used structured methods in the field of multi-omics integration which have been shown to perform well, most notably for clustering tasks [13, 14]: Joint Dimension Reduction (JDR) methods. This class of methods aims to estimate a low-dimensional (latent) space subsuming the main information contained in a multi-omics dataset by exploiting the joint signal among the different omic layers. Furthermore, JDR methods are easy to interpret as graphical representations of the position of individuals in the latent space can allow to understand the captured information. Moreover, weights associated to each predictor characterize their importance in the computation of this space.

As shown by Cantini et al. [14], latent spaces of JDR methods are significantly linked to survival data. Here, we quantify the predictive power for survival of such latent spaces for new patients, which has not been done in this previous work. Dimension reduction methods are however traditionally unsupervised. To tackle this limitation it has been proposed to use them as a first step before a predictive task [15, 16, 17, 18]. Here, we apply this approach to survival prediction, following Chegraoui et al. [18], by using a Cox regression on the latent space determined by joint dimension reduction. Then, being able to predict the survival of new patients requires to position new individuals in previously determined latent spaces, which involved methodological work for some JDR methods. With this approach, we benchmark 12 JDR methods on the 18 TCGA cancer datasets used by Herrmann et al. [7], as well as the best performing methods in this previous benchmark, against a similar baseline of a Cox model estimated only on clinical variables. In addition to prediction from both clinical and omics data, we also consider the case where prediction is drawn from omics data alone. Furthermore, we demonstrate that results from JDR methods are interpretable: using a bootstrap approach, we show that they can be used to identify reliable prognostic biomarkers. Further analysis of these multi-omics markers in the case of Low-Grade Glioma leads to clinically-relevant interpretations, supported by the current literature.

# 2 Results

## 2.1 Principles of tested JDR methods

In this article, $x$ denotes a scalar, $\mathbf{x}$ a vector, and $\mathbf{X}$ a matrix. In the case of multi-omics data, each type of omics data is expressed under the form of its own matrix $\mathbf{X}_k$ of size $n \times p_k$, $k \in \{1, \dots, K\}$. The number of individuals is $n$ while $p_k$ is the number of variables for omic $k$. $\mathbf{X}_k$ will be designated as blocks, or (omics) modalities.

Joint Dimension Reduction (JDR) methods are multiblock-specific Dimension Reduction (DR) methods, i.e., they compute a latent space of small dimension, common to all blocks considered. The JDR methods we evaluated can be categorized based on their dimension reduction strategy: covariance-based (Cov) methods, matrix factorization (MF) methods [13] and tensor factorization (TF) methods [14].

Matrix factorization methods seek an approximation of $\mathbf{X}_k$ of the form $\mathbf{Z}\mathbf{W}_k^\top$, with $\mathbf{Z} \in \mathbb{R}^{n \times R}$ a matrix describing the position of each individual in a latent space of dimension $R$ common to all modalities, and $\mathbf{W}_k \in \mathbb{R}^{p_k \times R}$ a weight matrix specific to each block allowing to reconstruct $\mathbf{X}_k$ from $\mathbf{Z}$ (see figure 1A). Among matrix factorization methods, we considered Joint and Individual Variations Explained (JIVE) [19], which considers both a latent space common to all modalities and distinct latent spaces specific to each block, Integrative Non-negative Matrix Factorization (IntNMF) [20], that includes non-negativity constraints on the weights and latent spaces, leading to good clustering properties, iCluster, another method designed for clustering that uses maximum likelihood, [21, 22] and Multi-Omics Factor Analysis (MOFA) [23, 24], a Bayesian method able to handle multiple data distributions.

Covariance-based methods seek, for each block, a weights vector $\mathbf{w}_k$ and a component (latent variable) $\mathbf{z}_k = \mathbf{X}_k\mathbf{w}_k$ that maximizes the covariance (or correlation) between block components (see figure 1B). In a procedure called deflation, this maximization is repeated $R$ times, each time after removing the effects of the already determined $\mathbf{z}_k^r$ from the corresponding block. This gives us $R \times K$ $\mathbf{w}_k^r$ with associated $\mathbf{z}_k^r$ which forms the latent space $\mathbf{Z}$. Of the covariance-based methods, we considered Regularized Generalized Canonical Correlation Analysis (RGCCA) [25, 26] which generalizes Canonical Correlation Analysis [27] and Partial Least Squares [28] to more than 2 blocks by maximizing the covariance between user-determined pairs of block components, Sparse Generalized Canonical Correlation Analysis (SGCCA) [29], a variant of RGCCA constraining the $\mathbf{w}_k^r$ to sparsity, and Multiple Co-Inertia Analysis (MCIA) [30], which maximizes covariance between each block component and a synthetic center, while using the same preprocessing steps as Non-Symmetric Correspondence Analysis (NSC) [31].

Tensor factorization methods require the omics to be put in the form of a tensor, in this case a 3-way table $\boldsymbol{\mathcal{X}} \in \mathbb{R}^{n \times p \times K}$ with entries for each combination of individual, variable (e.g., gene), and omics. This requires each dataset to have the same set of features, thus we aggregated each omics modality at the gene level (see Methods 4.2 for details). Among tensor factorization methods, we considered tensor Principal Component Analysis (tPCA) [32], an extension of PCA to tensor data, which seeks three semi-orthogonal matrices: the latent space $\mathbf{Z} \in \mathbb{R}^{n \times R}$, the weight matrix $\mathbf{W} \in \mathbb{R}^{p \times R}$ common to all omics, and $\mathbf{O} \in \mathbb{R}^{K \times R}$ where each element describes the contribution of omic modality $k$ to component $r$. Blocks $\mathbf{X}_k$ can then be approximated by $\mathbf{Z}\boldsymbol{\Delta}_k\mathbf{W}^\top$, where $\boldsymbol{\Delta}_k$ is a diagonal matrix containing the $k$-th line of $\mathbf{O}$. This approach further augments the compression of the data by using $R(n+p+K)$ parameters to reconstruct the data, instead of $R(n+Kp)$ for a non-tensor matrix factorization, enhancing the noise reduction properties of dimension reduction. Such model will focus on extracting molecular signatures where genes are jointly impacted across omics, enhancing the model interpretation. In addition to tPCA, we considered tensor Independent Component Analysis (tICA), which computes rotations such that the tPCA components are not only uncorrelated but also independent [33]. Detailed equations behind each JDR method evoked until now are available in the Supplementary Methods.

While JDR methods are typically unsupervised, for RGCCA, SGCCA, and JIVE, it is possible to supervise the construction of the latent space using survival information. For RGCCA and SGCCA, supervision is achieved by maximizing covariance between each block and a response block $\mathbf{y}$. In the case of survival prediction, $\mathbf{y}$ is the deviance residual of the null (without variables) Cox model [34, 18, 35]. Supervised JIVE (sJIVE) [36] proposes to minimize a convex combination of the original JIVE matrix decomposition problem and a least-squares criterion corresponding to the linear regression of both common and block-specific latent spaces on $\mathbf{y}$. The response variable we used is the same deviance residuals used for RGCCA and SGCCA. The obtained latent space is thus intended to capture more information related to survival.

Alongside JDR methods, we also considered two kinds of PCA as a control to compare classical Dimension Reduction (DR) methods against JDR ones. The first is a PCA in which all modalities are concatenated before applying PCA, corresponding to Early Integration (EI) [10]. In the second one, PCA is applied to all modalities independently and latent spaces are concatenated, corresponding to Late Integration (LI), while most JDR methods would be classified as Intermediate Integration.

We also considered a subset of the methods benchmarked by Herrmann et al. [7]: LASSO and survival random forests (ranger) [37, 38] among methods they classified as naive, i.e., they do not consider the block structure of multi-omics data, and IPF-LASSO, Priority-LASSO, Priority-LASSO clinical favoring and BlockForest among structured methods. Finally, to compare the performance of JDR methods against a baseline, we used the same references as in this previous benchmark [7]: a Cox model using only clinical data, and the Kaplan-Meier estimator, a non-parametric estimator for survival curves. All methods tested here are available as R packages: see table 1 for packages and functions used.

## 2.2 Survival Prediction from JDR methods

JDR methods are traditionally unsupervised methods that are not specifically designed for predictive tasks. This has already been tackled in the literature by using the latent space produced by Dimension Reduction methods as an input to a supervised model [15, 17, 16, 18], an approach we call sequential modeling. Here, to predict survival, we use the Cox model, a semi-parametric linear model for censored time-to-event data. Unlike in Herrmann et al. [7] where clinical data are always considered alongside omics data, we here also predict survival without clinical data, to assess the predictive capabilities of omics data alone. When considered, clinical data are added to the latent space when fitting the Cox model, to mimic the clinical favoring approaches that showed good performance in Herrmann et al. [7]. To alleviate issues arising from the multicolinearity in the latent space of covariance-based methods, we use a $\ell_2$-regularized Cox model [39] for all methods. To control for the impact of regularization on our baseline, we added an $\ell_2$-regularized Cox model using only clinical variables to the reference methods.

To benchmark this sequential modeling approach, we need to predict the position, in the estimated latent space, of each individual of a test set, without retraining the model. This is not a standard task for most JDR methods, and while RGCCA, JIVE, sJIVE and tPCA (+tFOBI) provided predictions, we had to develop prediction methods for IntNMF, iCluster and MCIA. For MOFA, we used MOTL [40], a transfer learning approach specifically developed for this method (see Methods 4.3 and Supplementary Methods for more details).

We run this sequential modeling approach inside a nested Cross-Validation (CV) loop, where the inner layer is a 10-fold CV used to select the best hyperparameters, most notably the number of latent variables and the regularization parameter of the Cox model (see figure 1C for a diagram of nested cross-validation). The outer layer is a repeated $5 \times 10$-fold CV and is used to measure survival prediction performance through two metrics: Uno's C-Index [41], a discrimination metric, and the Integrated Brier Score (IBS) [42], which is close to a prediction error (see Methods 4.4, 4.5 for more details on the CV procedure and JDR methods' hyperparameters; hyperparameters for methods from Herrmann et al. [7] were tuned as described in their study).

This CV setup and all subsequent experiments were run on the same 18 TCGA datasets as in Herrmann et al. [7], with 4 different omics modalities: intra-patient normalized gene (RNA) and micro-RNA (miRNA) expression, Copy Number Variants (CNV) and somatic mutations (mutations). The number of individuals per dataset ranges from 35 (LAML) to 735 (BRCA), with a total number of variables ranging from 90 151 to 103 366. The nested CV was run independently on each dataset using a framework similar to Herrmann et al. [7] and largely updated (see Methods 4.1 and 4.6 for details on datasets and on the computational framework).

## 2.3 JDR leads to reliable survival predictions across 18 TCGA cancers

Performances of each method obtained through this CV can be seen in figure 2. For each cancer, the 50 survival prediction performances are aggregated by taking the median over the resamplings. Thus, each boxplot is drawn from 18 values, one for each cancer data set (unaggregated boxplots, one for each cancer, can be seen in supplementary, figures S9 to S26). We observe no effect of the regularization on the baseline (non-regularized Cox model on Clinical data). Globally, with clinical data, we observe that JDR methods match the performance of the baseline in C-Index and can significantly outperform the performance of the baseline in IBS. The same observations can be made for PCAs. For structured and naive methods tested in [7], BlockForest, which has repeatedly been shown as the best performing method in this context [7, 10, 12, 43], is the only one that almost matches the reference model in median, though it does not lead to significantly better results. Without clinical data, performance decreases across all methods, both in C-Index and IBS, however most of the structured and JDR methods still match the performance of the baseline in IBS. Of important notice, BlockForest, sJIVE, S/RGCCA Sup. and SGCCA Unsup. are not significantly inferior to the baseline both in IBS and C-index.

A clear tendency towards a specific JDR strategy is hardly seen. Tensor Factorization methods does not lead to any significant results with clinical data nor do they match the baseline when assessed without clinical data. Furthermore, applying or not tensorial ICA (here with tFOBI) after tPCA does not seem to impact the results. For Covariance-Based methods, MCIA yields median performance results similar to S/RGCCA except for C-index with clinical data. The main difference among evaluated Cov-based methods lies in MCIA's preprocessing step originally introduced to analyze contingency tables. SGCCA's median performance results are close to its non-sparse counterpart, RGCCA, though it seems to be slightly lower, except for the supervised version in C-index with clinical data. This tendency is in agreement with the literature on $\ell_1$ vs. $\ell_2$ regularization. For MF methods, MOFA and iCluster lead systematically to lower median performance results than JIVE/sJIVE. This is also the case for intNMF but for C-index only. The systematic difference between these models is the ability of JIVE/sJIVE to extract specific omic effects along with common ones, the latter being the only focus of intNMF/MOFA/iCluster. Looking at the difference in performance between PCAs, these specific effects seem indeed important as LI-PCA systematically overcome median performance results of EI-PCA, except for IBS without clinical data where they match. This superior tendency of LI over EI has been observed multiple times in the literature [11, 10].

In the end, across all compared methods, LI-PCA and JIVE and S/RGCCA families consistently show good and similar performance results, either matching or exceeding the baseline in most cases. These observations are consistent with the benchmark from Cantini et al. [14], which concludes that RGCCA, JIVE and MCIA (LI-PCA not tested) are the best methods for survival prediction, though not evaluated through cross-validation.

The best non-baseline methods for each setting (combination of presence/absence of clinical data and metric, e.g., Uno's C-Index with clinical data) and cancer are presented in table 2. Each tested method is represented at least once as the best one. JDR methods are only slightly more represented in the best methods than non-JDR ones. Thus, no method can be excluded if searching for best performing method. The only cancer where performance is degraded by using omics data in conjunction with clinical data is UCEC, both in C-Index and IBS. We observe improvement above 10% in 19/72 combinations of settings and cancer, covering 7/18 cancers. LAML is the only where all 4 tested settings are concerned, with improvement ranging from 21% up to 28% in C-Index with and without clinical data respectively. COAD and SARC present improvements in 3/4 settings. HNSC/KIRP and LGG/LIHC present improvements above 10% (even 30% for KIRP) both with and without clinical data for C-index and IBS respectively. Of these improvements, 11 cases are due to JDR methods, 3 to non-JDR structured methods, and 5 to naive methods.

We present a heatmap of relative difference between each method and the baseline for each cancer, metrics and presence/absence of clinical data in figure 3. Hierarchical clusterings were computed for cancers and methods independently in each case. We notice a clear effect of the cancers on relative difference, dominating the method effect. With clinical data, for both metrics, most methods improve performances, relative to the baseline, for LGG, SARC, HNSC, COAD and LAML, while improvements are only seen in C-index for KIRP and in IBS for LIHC/PAAD. For all other cancers, performances are equivalent to the baseline. Without clinical data, for both metrics, most methods improve performances for HNSC, COAD and KIRP, while only in IBS for LGG/SARC/LIHC/LAML. Furthermore, for both metrics again, prognosis precision mostly decreases for SKCM, KIRC, LUAD, STAD and UCEC, while only in C-index for all the other cancers but LAML/LUSC/LGG. All other cancers not mentioned in IBS are associated with prognosis equivalent to the baseline. In the end, across most methods, improvements are observed for all conditions for HNSC and COAD, while for 3/4 conditions for LGG/SARC/KIRP/LAML. These datasets have very different sizes, with LAML being the smallest dataset (35 patients), and LGG and HNSC being among the largest datasets (419 and 443 patients, respectively). Decrease in prognosis precision in only seen without clinical data and for both metrics in only 5 cancers (ranging from 249 to 426 patients). Although sample size does not seem to impact the relative difference between each method and the baseline, it affects the absolute performance of methods (statistically significant only for IBS, see Supplementary Results 2.1).

As for methods, it seems that covariance-based methods have a slightly better performance on COAD than other types of methods. LASSO and, to a lesser extent ranger also seems to have a performance lower than other methods. This is in line with conclusions from Herrmann et al. [7] that structured methods perform better than naive ones. Interestingly, MOFA, iCluster and IntNMF are clustered together in all 4 settings, while also with EI-PCA in all settings but IBS without clinical data. As these methods only extract effects common to all omics, it could explain the similarity of their performances across cancers.

Technical information about the methods are available in table 1 for error rates and in Supplementary Results 2.2 for software availability and running time. Notably, JDR methods are slower than most methods from Herrmann et al. [7]'s study (except for BlockForest), which are themselves slower than reference methods (see figure S8). Among JDR methods, matrix and tensor factorization methods tend to be slower than covariance-based methods. Both PCA have however a running time comparable to methods from the previous study.

## 2.4 Omics data contribute to the prediction for JDR methods

To estimate the part of predictions coming from omics or clinical data, for each cancer/outer fold/JDR method, an omics contribution is computed (value between 0 and 1) from the corresponding Cox model. Contributions are averaged across outer folds and displayed in figure 4 (see Methods 4.7 for more details). We observe on the boxplot that for most JDR methods, as well as for LI-PCA, around half of the prediction comes from omics data. However, for iCluster, MOFA and IntNMF, omics data contributed less to predictions ($\sim$25%). Apart from being all MF models, they are also the only JDR methods focusing strictly on common effects, which might explain these results. This is also the case for the DR model EI-PCA using less omics data than its counterpart focusing only on specific effects, LI-PCA. Interestingly, EI-PCA's omic latent space seems to contribute more to survival predictions than iCluster/MOFA/IntNMF, which might be explained by PCA's specific orthogonality constraints on the latent space. On the contrary, supervised versions of S/RGCCA use omics data nearly exclusively for predictions. sJIVE behaves differently, as the contribution of omics data to predictions is similar to its unsupervised version, JIVE. It might be explained by differences in supervision strategies (regression vs. covariance).

We can see on the heatmap of omics contribution per cancer and method an important variability between cancers, with most methods (except for EI-PCA, IntNMF, iCluster and MOFA) using more omics data on BRCA, HNSC, COAD, KIRP and LIHC and less on STAD, ESCA and LUAD. S/RGCCA Sup. are the only methods not following this trend, but use less omics data on PAAD and KIRC. In addition to being cancers with a high use of omics data, HNSC, COAD and KIRP are also cancers where performance is consistently improved over the baseline, both with and without clinical data (see figure 3), and so may be cancers where omics data carry relevant information for prognosis that both partly overlap with clinical data and complement them. As opposed to STAD and LUAD where predictions are systematically decreased without clinical data compared to the baseline while similar to it with clinical data.

## 2.5 Bootstrap of JDR methods identify prognostic biomarkers

To demonstrate the interpretability of dimension reduction methods, we ran bootstrap experiments to estimate, for each method, confidence intervals of variables' weights with regard to the JDR model, as well as latent components weights in the Cox model. We selected RGCCA both supervised and unsupervised, JIVE, sJIVE, and the LI-PCA on all cancers to run these experiments, as they are methods with a consistent good performance. SGCCA was dismissed for this analysis as sparsity induce additional instability across bootstrap samples, which have to be investigated with dedicated procedures. Hence, for each cancer, each model was fit on 99 resamplings with replacement to compute confidence intervals and corresponding p-values for each variable and component (see Methods 4.7 for details). The proportion of latent variables that are significant for the Cox model as well as the proportion of significant genes per model can be seen in figure 5. We computed the proportion of significant variables, for each block and each cancer. For JDR models, when there are multiple components for a block, we computed the mean proportion of significant variables across components associated to this block. Finally, we represent the mean and 95% normal confidence interval for each method and each block. For Cox models, we observe that when clinical data are added to the latent space, a consistent proportion (between 14.8 % and 18.1 %) of clinical variables are significant predictors of survival. Out of these clinical variables, the most retrieved one is the age at diagnosis, being found in 30 Cox models out of 18 cancers $\times$ 5 JDR methods $=$ 90 Cox models, as this variable is present for all 18 cancers. Cancer stages may be even more systematically found significantly linked to survival (see table S8), though varying in presence and denomination across cancers. Interestingly, both SKCM and UCEC have a clinical variable significant for all 5 models: the Breslow thickness (melanoma's thickness) and the percentage of myometrial (muscular layer of the uterine wall) invasion by the tumor. Both seems to bear information at a macroscopic scale, harder to grasp directly with molecular data, though very relevant for prognosis. This might explain why, for SKCM and UCEC, predictions systematically dropped for all models and metrics without clinical data.

Components computed from omics data are used less consistently between methods and with a lower overall proportion for all JDR methods except for supervised RGCCA, where a high proportion of omics components are significant in Cox models (proportions between 76.7 % and 93.3 %) in coherence with figure 4. Regarding JDR models, significant genes are found in all blocks both with and without clinical data. LI-PCA shows a higher overall proportion of significant variables, while RGCCA shows a lower proportion. For most methods, the blocks with the highest proportions of significant genes are the RNA and miRNA blocks. In all methods except JIVE with clinical data, the proportion of significant variables is lower in the mutation block than in others. JIVE, both with and without clinical data, shows a high variance of significant mutations.

## 2.6 JDR models interpretation leads to multi-omic signatures linked to survival in LGG

We then selected components that are significant for the Cox model and performed gene set enrichment analysis (GSEA) on the bootstrap ratios (see Methods 4.7) associated with these components. We used MSigDB Hallmarks as gene sets for RNA components [44] and TAM 2.0 functions set for miRNA components [45]. For reasons of brevity, we chose to show the results of 4 different models: JIVE with clinical data (JIVE clin.), as well as JIVE, RGCCA Sup. and RGCCA Unsup. without clinical data, on Low Grade Glioma (LGG). This dataset was chosen for multiple reasons: (i) RGCCA Sup., RGCCA Unsup., LI-PCA, JIVE and sJIVE outperform the baseline in IBS without clinical data (ii) all bootstrapped methods use omics data to a high degree to predict survival for LGG when clinical data are present (figure 4). In the cases of RGCCA models, the hyperparameters with and without clinical data were identical (except for the regularization parameter) so we favored models without clinical data for interpretation (see tables S3 and S4).

For each of these models, the top 10 significantly enriched gene sets (ranked by gene ratio, Q-value threshold of 0.05) associated with each significant RNA and miRNA component according to bootstrapped Cox models can be seen in figure 6. In RNA components, 12 gene sets are identified by at least one JIVE and one RGCCA model, all with deleterious effects on survival. Out of these 12, 6 are categorized as immune gene sets: Allograft Rejection, Coagulation, IL6/Jak/STAT3 Signaling, Inflammatory Response, Interferon-$\alpha$ Response and Interferon-$\gamma$ Response, 1 as development: Epithelial Mesenchymal Transition, 1 as pathway: Apoptosis, 2 as proliferation: E2F Targets and G2M Checkpoint, and 2 as signaling: TNF-$\alpha$ Signaling via NF-$\kappa$B and mTORC1 Signaling. As such, most of the prognostic signal in transcriptomics data for LGG seems to come from immune processes, with deleterious overexpression for these hallmarks. Among these immune processes, interferon alpha and gamma responses are found enriched in all components and models, while IL6/Jak/STAT3 Signaling and Allograft Rejection are enriched in all but one component. IL6 and STAT3 related genes expression has been shown to correlate with tumor grade and patients survival in gliomas [46, 47, 48] while the IL6/Jak/STAT3 pathway has been shown to promote tumor growth and immune evasion [49, 50]. Sustained interferon signaling can also promote tumor progression and immunosuppression. Notably, prolonged interferon-$\gamma$ signaling impacts STAT3 signaling, while prolonged interferon-$\alpha$ signaling is linked to higher IL6 expression [51, 49]. Mesenchymal transition is another hallmark identified by all but one components which is an important driver of survival as it is linked to recurrent and treatment-resistant tumors [52]. Of particular interest, the mesenchymal transition is affected by other enriched pathways, namely the interferon alpha response through NF-$\kappa$B as well as IL6/Jak/STAT3 signaling [49]. We can hypothesize that these components capture the signature of a glioma subtype exhibiting similar mechanisms to the mesenchymal subtype of glioblastoma, which is associated to a poor prognosis [53]. This is explored further in Supplementary Results 2.3 where we show that JIVE clin. retrieve 3 known subtypes of low grade gliomas [54], among which *IDH*-wildtype gliomas that are closer to glioblastoma [55, 56]. Finally, the E2F targets and G2M checkpoints hallmarks are involved in cell cycle control, a commonly dysregulated process in cancer [57].

In miRNA, 15 gene sets are found by both a JIVE and RGCCA model: Adipocyte Differentiation, Bone Re-generation, Cell Division, DNA Damage Response, Folliculogenesis, Glucose Metabolism, Granulopoiesis, Innate Immunity, Neuron Differentiation, Onco-MiRNAs, Osteoblast Differentiation, Oxydative Stress, Peritoneal Cavity Homeostasis, T-Cell Differentiation and Toxicity. All are associated to a risk increase, except for Neuron Differentiation which is associated to a risk decrease by most models. Granulopoiesis, T-Cell Differentiation and Innate Immunity are immune processes and may be linked to the immune signature we identified in RNA components. Cell Division and Glucose Metabolism may be linked to tumor proliferation, while the mostly protective effect of Neuron Differentiation could be linked to a subtype similar to the proneural subtype of glioblastoma, which has a better prognosis than other glioblastoma subtypes [53].

To further interpretation on Low Grade Glioma, we represented the bootstrap ratios of the CNV components of JIVE, JIVE clin., RGCCA Sup. and RGCCA Unsup. which can be seen in figure 7. All components are significant for survival prediction, except for the cnv.1 component of RGCCA Unsup., which has still been considered because it is by design highly correlated with its corresponding RNA and miRNA components. All models identify strong signals on the short arm of chromosome 1 and the long arm of chromosome 19, corresponding to the frequent 1p/19q codeletion identified in the literature [54]. In most components this is a protective signal, in agreement with Wen et al. [54]. In the third common component of JIVE clin. the effect attributed to this codeletion is deleterious, contrary to all other components as well as the literature. This effect is explored in Supplementary Results 2.3 and may be explained by the manner in which each JIVE clin. components differentiate between subtypes. Another event observed in the literature to which we can associate a signal retrieved by all components is the deleterious loss of chromosome 10 which often co-occurs with a chromosome 7 gain [55]. A final signal of interest is the peak in chromosome 9p, which occurs in the region of genes *CDKN2A/B*, where a deletion have been associated to a poorer prognosis [58], in agreement with the signal observed here.

For JIVE, JIVE clin., RGCCA Sup. and RGCCA Unsup., the bootstrap ratios of their mutation components is represented in figure 8. Like for CNVs, all components are significative, except for the mutation.1 component of RGCCA Unsup. The JIVE mutation.1 component does not identify any significant gene. All common components of JIVE clin. identify *IDH1* as protective, the first two common components of JIVE clin. identify *FUPB1* and *CIC* as protective, while the second component identifies *NOTCH1* and *IDH2* mutations as protective, in accordance with type I glioma defined in Wen et al. [54]. For deleterious mutated genes, all three components identify *EGFR*, while components 2 and 3 identify *PTEN* and component 3 *NF1*, all in agreement with type III glioma [54]. Markers of type II glioma, namely *TP53* and *ATRX*, are identified with opposite effects in component 2 and 3. *CIC* and *FUBP1* are other genes identified with opposite effects in these two components. These sign changes are coherent with subtypes identified by JIVE clin. and explored in Supplementary 2.3. RGCCA Unsup. identifies *FUBP1*, *IDH1*, *EGFR*, *NOTCH1*, *NF1* and *CIC* with coherent effects. Other identified genes have no known prognostic mutations in the glioma literature: *TTN* and *FAT2* for the second common component of JIVE clin., and *PIK3CA* for the third, as well as 11 genes identified by RGCCA Unsup. RGCCA Sup. identifies a high number of genes as significant (43), most of those unknown in the literature. They must be interpreted with caution, as comparing an RGCCA Sup. model fit only on the mutation block to a model fit on all blocks reveals that RGCCA Sup. is overfit on the mutation block (Supplementary figure S59). Accordingly, the mutation block contributes the most to prediction in this case (supplementary figure S55; see Methods 4.7 for details on how this figure was constructed). We hypothesize that RGCCA Sup. uses rare mutations to distinguish individuals with specific survival times. Still, *IDH1/2*, *CIC* and *EGFR* are identified with effects coherent with the literature.

# 3 Discussion

In this study, we rigorously benchmarked 12 JDR methods in the context of survival prediction. This class of methods is widely used in the field of multi-omics study and was not represented in Herrmann et al. [7]'s benchmark. By using the same datasets and a framework similar to this previous study, we allow for an easier comparison of results. Moreover, we improved on the work of Cantini et al. [14] by testing JDR in a predictive task, most notably by assessing predictive power through survival-specific metrics on individuals unseen by the model, through a cross-validation procedure. In order to do this, we had to develop specific prediction methods for some of the tested algorithms, an aspect often unexplored in JDR methods which focus on unsupervised analysis. Thanks to this predictive focus, our work is relevant both for identifying new biomarkers in a research cohort, and prognostic testing in clinical routine. While we focus on survival prediction, our work can be transposed to other predictive tasks, through both the novel prediction methods evoked above, and the flexibility of sequential modeling.

From this benchmark, it appears that using JDR methods along sequential modeling to combine omics and clinical data improve on average survival prediction performances over a baseline using only clinical data. This is not true for methods tested by Herrmann et al. [7] (structured or not), in agreement with their original study, even though our aggregation strategy differs. Without clinical data, BlockForest along with RGCCA and JIVE models are the only methods able to match the predictive performance of clinical data for both metrics. This is particularly promising for the discovery of new biomarkers. It appears however that no single method is the best for all cancers and that no method (JDR or not, structured or not) can be excluded *a priori* when looking for the best performances for a dedicated cancer. We also notice a strong effect of cancers on relative performance between each method and the baseline, with improvements/decreases spanning from the smallest to the highest sample sizes available. We hypothesize that, for some cancers, the link between omics data and survival is stronger, or at least easier to extract using the methods presented here. For instance, a cancer may present subtypes with clear survival differences that can be captured by a given model, as seen with low grade glioma.

Part of performance gain of JDR methods may come from the dimension reduction rather than the joint aspect as LI-PCA does not consider links between omics, and outperforms some of them. Furthermore, as LI-PCA also outperforms EI-PCA, it seems that extracting only omic-specific effects is a superior strategy to extracting only joint ones. In the sequential modeling framework, extracting omic-specific effects prior to fitting a Cox model allows this survival model to decipher which omic is involved and to what extent. It resonates with the work of Wissel et al. [12] arguing that, in a similar context, the best strategy to perform multi-omics integration is by performing omic selection in regard of the survival task.

Although LI-PCA shows performance similar to some JDR methods, the most consistently well-performing methods are JIVE and S/RGCCA, which we recommend among the methods we tested. This is consistent with the work of Cantini et al. [14]. As pointed by their study, these methods have in common that they compute latent variables specific to each modality thus relaxing the constraint of most JDR methods to have only components common to all blocks. However their theoretical approaches are different: on one hand, JIVE seeks both common and specific effects unique to each block, uncorrelated with each other. On the other hand RGCCA, by maximizing covariances, looks for components that both explain part of their block's variance as well as correlate with their connected blocks. Hence a component can carry both common and specific effects. Their flexibility in extracting both types of effects may explain their performance in regard to LI-PCA. In addition to components common or specific to each modality, some methods consider components linked to a strict subset of omics modalities (e.g. a component computed from RNA and miRNA only) [59, 60]. This type of methods is however still uncommon in the current state of the art and computationally costly, as they have to consider the whole space of possible combinations of omics modalities.

Supervision of JDR methods can be a useful tool for interpretation by reinforcing the survival signal in the latent space, as shown with the supervision of RGCCA producing a higher proportion of latent variables significantly linked to survival. This is however not reproduced with the supervised version of JIVE, which may be due either to a normalization that differs from regular JIVE, or a less strict convergence threshold than JIVE. For RGCCA specifically, the supervised version matches the performance of the unsupervised one while requiring to compute fewer components and converging quicker, providing an advantage in terms of computing time. However, as seen in the case of LGG, RGCCA Sup. may be prone to overfit at least on some blocks. This undermine some of the advantages of supervision, as overfit models, being less generalizable, have degraded performance and are harder to interpret. While the deviance residuals are an easy and efficient way to supervise dimension reduction methods for survival data [34, 35], this approach relies on an approximation of the Cox likelihood. Hence supervised dimension reduction methods could benefit from an exact approach using either the Cox partial likelihood or another survival-specific metric.

For some cancers, JDR methods exploit well omics data to predict survival, allowing them to match or outperform the baseline model with or without clinical data. This indicates that further interpretation of how these results are obtained can help identify prognosis biomarkers. We showed that bootstrap is in this context a useful tool, allowing to quantify both the importance of each latent variable for survival prediction, and of individual genes in latent variables. Enrichment analysis can also be used to identify dysregulated pathways and mechanisms. In the specific case of low grade glioma, we were able to identify chromosomal events and mutated genes already known in the literature as linked to prognosis. We also identified sets of genes (hallmarks) whose expression is associated with a poorer prognosis that seems coherent with known cancer mechanisms. Still, these results must be interpreted with caution, as not every gene contributing to a component predictive of survival is itself associated with survival but rather to potential disease subtypes, as seen with low grade glioma. In practice, we recommend comparing the results of at least two models using different integration strategies to draw clearer interpretations. For low grade glioma, JIVE clin. identifies the clearest signatures among the tested models. For this cancer, cross-validation selected only common components for JIVE clin. As all omics contribute to common components, it may explain the easier interpretability of such components. This strengthens the idea that JDR able to extract both omic-specific and joint effects should be preferred as such flexibility improves prediction performances while joint components enhance interpretability.

# 4 Methods

## 4.1 Datasets

To evaluate the prognostic performance of joint dimension reduction methods, we used the same 18 TCGA datasets as Herrmann et al. [7], to allow for an easier comparison between studies. For each of these datasets, four molecular data types are available: mRNA and miRNA expression, Copy Number Variants per gene (CNV), all three in the tumour samples, as well as somatic mutations at the gene level. Herrmann et al. [7]'s authors downloaded data from TCGA using the R package TCGAbiolinks (function `GDCquery`) before release 32 (Gencode v22). For each cancer type, raw mRNA counts (`data.category = "Transcriptome Profiling", data.type = "Gene Expression Quantification", workflow.type = "HTSeq - Counts"`) were filtered out to remove low-expression genes ($< 0.5$ reads per million in $\geq 95\%$ of patients). Then log2-Counts Per Million (log2-CPM) were obtained using the `voom` function from the limma package. For miRNA, reads-per-millions value were selected (`data.category = "Transcriptome Profiling", data.type = "miRNA Expression Quantification"`) and similarly to mRNA, low-expression miRNAs were removed ($< 0.5$ reads per million in $\geq 95\%$ of patients) and log2-CPM were obtained with the `voom` function. For CNV, segment mean for all human Ensemble IDs were retrieved (`data.category = "Copy Number Variation", data.type = "Copy Number Segment"`) and null values across all patients were filtered out. For mutations (function `GDCquery_Maf` this time, with `pipelines = "mutect2"`), only genes with at least one somatic mutation in one patient were kept. Each gene value was set to 1 if it bears at least one somatic mutation and 0 otherwise. A varying number of clinical data are also available for each cancer, e.g., age, sex, and cancer stage for each patient. The number of individuals per dataset ranges from 35 (LAML) to 735 (BRCA), while the total number of variables is comprised between 90 151 and 103 366, of which roughly 60% are from the CNV block, 25% from RNA and 15% from mutations, while miRNA and clinical data represent less than 1% of the total number of variables (see supplementary table S2). Not a single value is missing across the 18 datasets.

## 4.2 Tensor construction for Tensor Factorization methods

As mentioned in the results section, tensor factorization methods require each modality to have the same set of features. Here, the RNA, mutation, and CNV datasets are already aggregated at the gene level, and as such we keep the features pertaining to the intersection of these three modalities. As the mutation block can be rather small (e.g. 2176 genes for LAML, see table S2) due to Herrmann et al. [7] listing only genes where somatic mutations are observed for at least one patient in a cancer type, we replaced genes present in the RNA dataset and missing in the mutation dataset by columns of 0, to explicitly take into account the absence of somatic mutation for them and in the end keep the maximum number of genes to build the tensor. Without this imputation, we would have lost many genes, as the mutation block is typically the smallest. We could have replaced genes present in the CNV dataset and missing in the mutation dataset by columns of 0. However, adding columns to the mutations based on the CNVs would have necessitated adding corresponding columns in the RNA blocks, as many genes in the CNV blocks are non-coding. Imputation is however more complex for the RNA blocks and would have required strong hypotheses on the expression of these non-coding genes. miRNAs remain to be aggregated at the gene level. Mature miRNAs IDs were retrieved from premature IDs using the R package miRBaseConverter [61]. Based on the approach from MONTI [62], we computed for each gene the mean expression value of the miRNAs targeting this gene. Validated targets were retrieved with the multimir R package [63]. As a premature miRNA can lead to multiple mature miRNAs, its set of targets was defined as the union of the sets of targets of its corresponding mature miRNAs.

## 4.3 Sequential Cox modeling

Nearly all JDR methods are unsupervised methods not designed for predictive tasks. In the literature, this problem has been tackled by using the latent space determined by a JDR method as variables of a predictive task, for example for JIVE in the context of survival prediction [16] and for R/SGCCA in the context of classification [29] or prognosis [18]. To predict survival from the latent space of JDR methods $\mathbf{Z}$, we use the same approach, which we call sequential modeling: we use our latent spaces as variables for a regularized Cox regression [2], for both supervised and unsupervised JDR methods. The Cox model is a semi-parametric linear model for time-to-event censored data. We selected the Cox model as it is one of the simplest models for censored data, furthermore it enhances the comparison to the reference model. Herrmann et al. [7] only considered methods using clinical data as well as omic data, in order to determine if omic data can improve prognostic testing. As we are also interested in how the latent space relates to survival data without consideration of clinical variables, we considered two possibilities for this sequential modeling approach: with and without adding clinical variables to the latent space. When applicable, the clinical data are added at the level of the Cox regression and not in the JDR method, to mimic the clinical favoring approaches that showed good performance in Herrmann et al. [7].

To benchmark this sequential modeling approach we need to make predictions for individuals on a test set, without retraining the model. For most JDR methods this is not a standard task, as they are unsupervised methods. For PCA, prediction for new individuals is straightforward: thanks to the orthogonality constraint, the weight matrix is a projection matrix that allow us to easily position new observations in the latent space determined during training. However, for most JDR methods presented, this orthogonality constraint is not valid. When possible, we used functions provided by packages (RGCCA, JIVE, sJIVE, tPCA and tPCA+tFOBI). When not possible (IntNMF, iCluster, MOFA, MCIA), we devised a prediction method to determine the position of the new individuals with the weights of the variables fixed, while staying as close as possible to the original optimization problem. For IntNMF, we used an internal function of the package to solve for the position of the new individuals. For iCluster we replicated in R the Expectation step of the original E-M algorithm in Fortran. We initially considered iClusterPlus [64] as it allows to take into account specific data distributions (binomial, multinomial, Gaussian and Poisson) as opposed to iCluster that only take into account Gaussian distributions. However we were not able to develop a working prediction function for this method: iClusterPlus uses a Metropolis-Hasting algorithm for the E step instead of the closed form solution of iCluster and we were not able to obtain convergence of this algorithm for new data. For MOFA we used an already published transfer learning approach, MOTL [40], which uses the same Bayesian framework as MOFA to compute the position of new individuals while fixing the weights. Finally, for MCIA we had to apply the Non-Symmetric Correspondence Analysis [31] pre-processing step on the new individuals, for which we tested two approaches and kept the better performing one. Prediction of the position of the new individuals for MCIA is then straightforward and similar to RGCCA. More details on these prediction methods are available in Supplementary Methods.

## 4.4 Cross-Validation setup for JDR methods

To both rigorously measure the performance of our sequential modeling approach and select the best hyperparameters (most notably the number of latent variables and the $\ell_2$ regularization parameter of the Cox model), we use a nested cross-validation approach, in a similar fashion to some methods benchmarked by Herrmann et al. [7], e.g. cv.glmnet (see figure 1C for a diagram of nested cross-validation). In the first layer, we use a 10-fold cross-validation to select the best set of hyperparameters, except for iCluster limited to 3-fold due to its high computationnal burder. The Cox partial log-likelihood was used as a metric for this first layer, determined with the approximation from Verweij et al. [65], which is the default approach of glmnet and can be seen as an approximation of the generalization error for the Cox likelihood. This first (inner) cross-validation is run inside each iteration of an outer repeated $10 \times 5$-fold cross-validation that is used to measure performance through the 2 survival metrics presented below. The repeated $10 \times 5$-fold CV for each cancer means that, taking into account all cancers, 18 cancer $\times$ 5 folds $\times$ 10 repeats = 900 folds are run for each model, and for each of these folds, 10 inner folds are run, multiplied by the number of combinations of hyperparameters tested.

We measure the performance in survival prediction of our models through 2 survival-specific metrics: Uno's C-Index [41] and the Integrated Brier Score (IBS) [42]. Intuitively, the C-Index compares the ranking of the individuals based on the risks predicted by the Cox model to the true ranking based on the actual time of death: it is roughly equivalent to the proportion of correctly ranked pairs of patients. The IBS computes for each patient the distance between the survival curve outputted by a model and an ideal survival curve which is a step functions whose value is 1 until observed time of death and 0 after. As such, it is more sensitive to small changes in prediction that do not affect ranking and should be preferred to C-Index as a metric to choose between models, however the latter is easier to interpret. Both of our survival metrics require estimating the censoring distribution and as such are highly sensitive to individuals censored at later times in the study [66, 67]. To mitigate this effect, both metrics are computed up to a time threshold which is the 0.9 quantile of censoring times following Wissel et al. [10]. In the end, for resamplings with execution errors, the performances are replaced with the median value of the resamplings for this cancer, when they represent less than 20% of the resamplings (10 resamplings). Otherwise, they are replaced by 0.5 for C-Index and 0.25 for IBS [7] which are values that represent the performance of a random prediction.

## 4.5 Considered Hyperparameters for JDR methods

The JDR methods we included have each their own set of hyperparameters. However, they all share two types of hyperpameters: the regularization strength of the Cox model due to our sequential modeling approach which we set using the same procedure as glmnet, and the number of latent variables, which is the only hyperparameter we tested for EI-PCA, LI-PCA, intNMF, MOFA and MCIA. We tested different numbers of latent variables in the set $\{1, 5, 10\}$ or $\{2, 5, 10\}$, the second case being for IntNMF, MOFA, MCIA and tFOBI which do not allow for only 1 latent variable. The limit of extracted components was set to 10 as we consider it a reasonable balance between extracted signal and computational cost. It is also the number of components used in the work of Cantini et al. [14]. Note that in the case of covariance-based methods, these numbers correspond to a number of latent variables per modality, which means that the latent space contains in practice four times (four omic blocks) as many variables. To determine the grid of regularization parameters, we run glmnet [39] on the whole outer training set in a process similar to cv.glmnet. The maximal value of the regularization parameter is thus the smallest value such that no variable is selected by the Cox model [68], and the minimum value is 100 times lower. We consider 100 values equally spaced on a log scale between these two values. Then, concerning JDR methods' hyperparameters, if some of them can be set to different values for each modality, we choose to fix these hyperparameters to be identical across all 4 modalities to reduce the computational cost of cross-validation. As such, for JIVE, we tested combinations of 0, 1, 5 or 10 common components with 0, 1, 5, or 10 individual components (without the case where no component at all is extracted, it leads to 15 sets of hyperparameters in total, regularization parameter excluded). For sJIVE, we added combinations with $\eta \in \{0.1, 0.5, 0.9\}$, modulating the amount of supervised learning (75 sets in total; see Supplementary Methods for more details on $\eta$). For RGCCA, we considered combinations of the number of latent variables ($\{1, 5, 10\}$) with various values of $\tau \in \{0.1, 1\}$, controlling the mix between correlation and covariance (6 sets; see Supplementary Methods for more details on $\tau$). For SGCCA, $\tau$ is set to 1 and the sparsity hyperparameter $s$ has to be tuned. We set $s$ at 25%, 50% or 75% of its maximal value, which depends on the number of variables in each block and leads to a non-sparse model (9 sets with the combinations of the number of latent variables). For iCluster, to limit the computational burden, we considered only two possible number of latent variables: either 1 or 5 and we set the sparsity hyperparameter to get roughly 0%, 33% or 66% of variables selected (6 sets). Theses values were determined empirically for each dataset by running iCluster on all samples. The frequency of selection of each hyperparameter combination for each model by the inner cross-validation loop can be seen in figure S27 to S40.

## 4.6 Benchmark implementation

All packages used are available on the CRAN or Bioconductor, except for sup.r.jive which contains sJIVE and is available under the form of a github repository. We had to modify two packages: IntNMF, to fix a bug affecting the choice of the best factorization when running multiple initializations, and iClusterPlus, to allow the Fortran code to compile in a conda environment. To use MOTL, we had to run MOFA directly from the mofapy2 Python package with reticulate as the R implementation does not save all parameters necessary for MOTL. iCluster requires a large quantity of RAM (around 100Gb per CPU) to run on the datasets. As few CPUs available on our cluster meet this requirement, we had to use 3 inner cross-validation folds (instead of 10) to be able to include this method, as mentioned above.

To implement this benchmark we used the mlr3 framekwork in R, as the mlr package used by Herrmann et al. [7] is now deprecated. As such, we developed an appropriate mlr3 learner for each method. These mlr3 learners implement the sequential modeling as well as the inner cross-validation. The outer cross-validation and metrics computation part is managed by mlr3. The learners for each JDR method are available in an R package. For BlockForest, IPF-LASSO and Priority-LASSO, we adapted learners already developed for mlr by Herrmann et al. [7]. The benchmark was run on a local computing cluster comprising a total of 2307 cores from various architectures. The R package batchtools was used to interface the mlr3 framekwork with the cluster. We developed a package to automate part of this process. See 5 for more details.

## 4.7 Interpretation

Omics contribution values are computed from the weight vectors $\boldsymbol{\beta}$ of the Cox models fitted on JDR methods latent space. More precisely, when considering clinical variables in addition to latent omics variables, we normalized $\boldsymbol{\beta}$: $\boldsymbol{\beta}_{\text{norm}} = \frac{\boldsymbol{\beta}}{\|\boldsymbol{\beta}\|_2}$ and with $S$ the set of latent omic variables, we computed $\sum_{i \in S} \beta^2_{\text{norm},i}$ for each fold of each model and cancer giving an estimation between 0 and 1 of how much the omics data contribute to the survival prediction. These values are then aggregated by taking the mean by cancer.

While bootstrapping JDR methods, it is important to consider each method invariances [34]. Indeed, JIVE and PCA show rotational invariance, meaning that all rotations of a solution of the JIVE/PCA optimization problem are equally good solutions. To mitigate this effect on the bootstrap samples, we realigned all latent spaces and weight matrices on a model fitted on the original dataset by determining for each bootstrap sample a Procrustes rotation between its weight matrices and the original weight matrices. RGCCA shows sign invariance, a special case of rotational invariance. We ensured that the correlation between the latent variables (and associated weights) of the bootstrapped model and the original model were positive by multiplying by -1 when necessary. LI-PCA, sJIVE and RGCCA remove null variance variables from the dataset: to take it into account, rotations and correlations were computed for each bootstrap sample considering only variables common to the bootstrapped model and the original model. In each resampling, weights of removed variables were set to 0. P-values were computed from the bootstrap samples using the bootstrap ratio, i.e. the ratio between the mean of a coefficient and its standard deviation [69]. Under the null hypothesis, bootstrap ratios follow a standard normal distribution. P-values were corrected using the Benjamini-Hochberg procedure. This procedure was applied separately twice: once for all p-values linked to JDR models (100 458 240 p-values in total), and once for all p-values linked to Cox models (4014 p-values in total).

Omics contribution were computed from the 99 bootstrap resamplings, in a similar manner to previously computed omics vs clinical data contribution, this time taking advantage of the realignment described above to compute contributions for each block. These contributions can be seen from figure S41 to S58.

Hyperparameters for the bootstrapped methods were chosen by selecting for each cancer the combination performing most often best in the inner cross-validation (see table S3-S7). When two or more hyperparameters combinations were tied, we selected the combination with the best median Cox log-likelihood (computed with the Verweij and Van Houwelingen approximation [65]).

For the enrichment analysis, we used Fast Gene Set Enrichment Analysis (FGSEA) [70] through the R package ClusterProfiler [71], ranking the variables using their bootstrap ratios, using as gene sets the hallmarks from MSigDB [44]. FDR was controlled by computing q-values [72] considering all pathways, all components, all methods and all cancers.

# 5 Code Availability

All scripts developed and used to run the benchmark are available on github at `https://github.com/vlegoff/MultiOBenchmark` along packages developed for this analysis. All methods and packages used for this benchmark have been installed in conda environments. The same github contains all conda configuration files (with a list of packages and their versions) and installation scripts necessary to create the environments.

# 6 Data Availability

All data used for this analysis and intermediary results produced are available on Zenodo at `https://zenodo.org/records/24402988`.

Table 1: **R packages and functions used to run each method**, along with the origin of the prediction function (custom or provided by the package), and the percentage of errors for each method. Number of calls corresponds to number of cancers (18) × number of outer folds (50) × number of hyperparameter combinations × presence/absence of clinical data (if relevant).

| Method name | Method type | Package::function | Prediction | Errors | Calls | Error rate |
|---|---|---|---|---|---|---|
| Kaplan-Meier [73] | Reference | survival::survfit | From package | 0 | 900 | 0.00% |
| Cox clin. [2] | | survival::coxph | | 0 | 900 | 0.00% |
| Cox L2 clin. [39] | | glmnet::glmnet | | 0 | 900 | 0.00% |
| LASSO [39] | Naive | | | 0 | 1,800 | 0.00% |
| ranger [37, 38] | | ranger::ranger | | 0 | 1,800 | 0.00% |
| IPF-LASSO [3, 74] | Structured | ipflasso::cvr.adaptive.ipflasso | | 420 | 1,800 | 23.33% |
| Priority-LASSO [4] | | prioritylasso::prioritylasso | | 74 | 1,800 | 4.11% |
| Priority-LASSO clin. fav. [4] | | | | 8 | 900 | 0.89% |
| BlockForest [6] | | blockForest::blockfor | | 0 | 1,800 | 0.00% |
| PCA Joint [75] | PCA | ade4::dudi.pca | | 0 | 5,400 | 0.00% |
| PCA Indiv. [75] | | | | 0 | 5,400 | 0.00% |
| JIVE [19] | JDR - MF | r.jive::jive | | 3 | 27,000 | 0.01% |
| sJIVE [36] | | sup.r.jive::sJIVE | | 5,109 | 81,000 | 6.31% |
| IntNMF [20] | | IntNMF::nmf.mnnals | Custom | 0 | 5,400 | 0.00% |
| iCluster [21, 22] | | iClusterPlus::iCluster2 | Custom | 1,022 | 10,800 | 9.46% |
| MOFA [23, 24] | | Python package mofapy2 | MOTL [40] | 0 | 5,400 | 0.00% |
| tPCA [32] | JDR - TF | tensorBSS::tPCA | From package | 0 | 5,400 | 0.00% |
| tPCA+tFOBI [33] | | tensorBSS::tFOBI | | 0 | 5,400 | 0.00% |
| RGCCA Unsup. [25, 26] | JDR - Cov | RGCCA::rgcca | | 693 | 10,800 | 6.42% |
| SGCCA Unsup. [29] | | | | 1,042 | 16,200 | 6.43% |
| RGCCA Sup. [25, 26, 34, 18] | | | | 665 | 10,800 | 6.16% |
| SGCCA Sup. [29, 34, 18] | | | | 1,088 | 16,200 | 6.72% |
| MCIA [30] | | omicade4::mcia | Custom | 10 | 5,400 | 0.19% |

Table 2: **Best method by cancer** Best methods and their median performance are given by cancer, for both metrics, with and without clinical data. Median performance of Cox clin. is given as reference. The name of the method is in bold if it is a JDR method. Cancers are sorted from the lowest number of patients (LAML) to the highest (BRCA). A + as exponent indicates that for the specific metric and setting, the best method also outperforms Cox clin.

| | Cox clin. | | With clinical data | | | | Without clinical data | | | |
|---|---|---|---|---|---|---|---|---|---|---|
| | C-Index | IBS | C-Index | | IBS | | C-Index | | IBS | |
| Cancer | Value | Value | Method | Value | Method | Value | Method | Value | Method | Value |
| LAML | 0.585 | 0.238 | BlockForest$^+$ | 0.754 | **iCluster** $^+$ | 0.179 | BlockForest$^+$ | 0.707 | ranger$^+$ | 0.185 |
| ESCA | 0.575 | 0.202 | **RGCCA Unsup.**$^+$ | 0.592 | IPF-LASSO$^+$ | 0.186 | **RGCCA Sup.** | 0.560 | **MOFA**+**MOTL**$^+$ & EI-PCA$^+$ | 0.193 |
| PAAD | 0.674 | 0.172 | **sJIVE**$^+$ | 0.711 | **sJIVE** $^+$ | 0.158 | **SGCCA Sup.** | 0.654 | EI-PCA$^+$ | 0.172 |
| SARC | 0.677 | 0.203 | **RGCCA Sup.** $^+$ | 0.752 | **RGCCA Sup.** $^+$ | 0.154 | LASSO$^+$ | 0.702 | **sJIVE** $^+$ | 0.168 |
| LIHC | 0.596 | 0.196 | **RGCCA Sup.**$^+$ | 0.614 | ranger$^+$ | 0.164 | **tPCA** | 0.588 | ranger$^+$ | 0.160 |
| KIRP | 0.526 | 0.133 | LI-PCA$^+$ | 0.693 | BlockForest$^+$ | 0.122 | **JIVE**$^+$ | 0.683 | **sJIVE** $^+$ | 0.123 |
| COAD | 0.620 | 0.142 | **SGCCA Sup.** $^+$ | 0.747 | **RGCCA Unsup.** $^+$ | 0.123 | **SGCCA Sup.** $^+$ | 0.712 | **SGCCA Unsup.** $^+$ | 0.121 |
| OV | 0.611 | 0.168 | Priority-LASSO$^+$ | 0.615 | **IntNMF**$^+$ | 0.165 | **IntNMF**$^+$ | 0.614 | **IntNMF**$^+$ | 0.165 |
| KIRC | 0.752 | 0.138 | LI-PCA$^+$ | 0.785 | EI-PCA$^+$ | 0.133 | **RGCCA Sup.** | 0.709 | EI-PCA | 0.149 |
| SKCM | 0.645 | 0.176 | LI-PCA$^+$ | 0.657 | EI-PCA$^+$ | 0.172 | ranger | 0.574 | LI-PCA | 0.192 |
| STAD | 0.657 | 0.179 | **SGCCA Unsup.**$^+$ | 0.660 | **IntNMF** $^+$ | 0.174 | **IntNMF** | 0.559 | **tPCA**+**tFOBI** | 0.194 |
| BLCA | 0.659 | 0.194 | Priority-LASSO$^+$ | 0.681 | **tPCA** $^+$ | 0.178 | **RGCCA Sup.** | 0.640 | **RGCCA Sup.**$^+$ | 0.191 |
| UCEC | 0.750 | 0.0903 | **IntNMF** | 0.741 | **IntNMF** | 0.0904 | LI-PCA | 0.677 | ranger | 0.0989 |
| LUSC | 0.551 | 0.207 | LI-PCA$^+$ | 0.566 | **sJIVE** $^+$ | 0.202 | EI-PCA$^+$ | 0.565 | **sJIVE**$^+$ | 0.204 |
| LGG | 0.733 | 0.124 | LASSO$^+$ | 0.792 | LASSO$^+$ | 0.100 | LASSO$^+$ | 0.749 | LASSO$^+$ | 0.110 |
| LUAD | 0.690 | 0.148 | Priority-LASSO$^+$ | 0.691 | Priority-LASSO clin. fav.$^+$ | 0.148 | **MCIA** | 0.570 | **MCIA** | 0.166 |
| HNSC | 0.545 | 0.203 | **RGCCA Sup.** $^+$ | 0.632 | **RGCCA Sup.** $^+$ | 0.188 | **SGCCA Sup.** $^+$ | 0.615 | **tPCA**+**tFOBI** $^+$ | 0.192 |
| BRCA | 0.670 | 0.0823 | **sJIVE** $^+$ | 0.708 | **sJIVE** $^+$ | 0.0774 | LI-PCA$^+$ | 0.675 | LI-PCA$^+$ | 0.0797 |

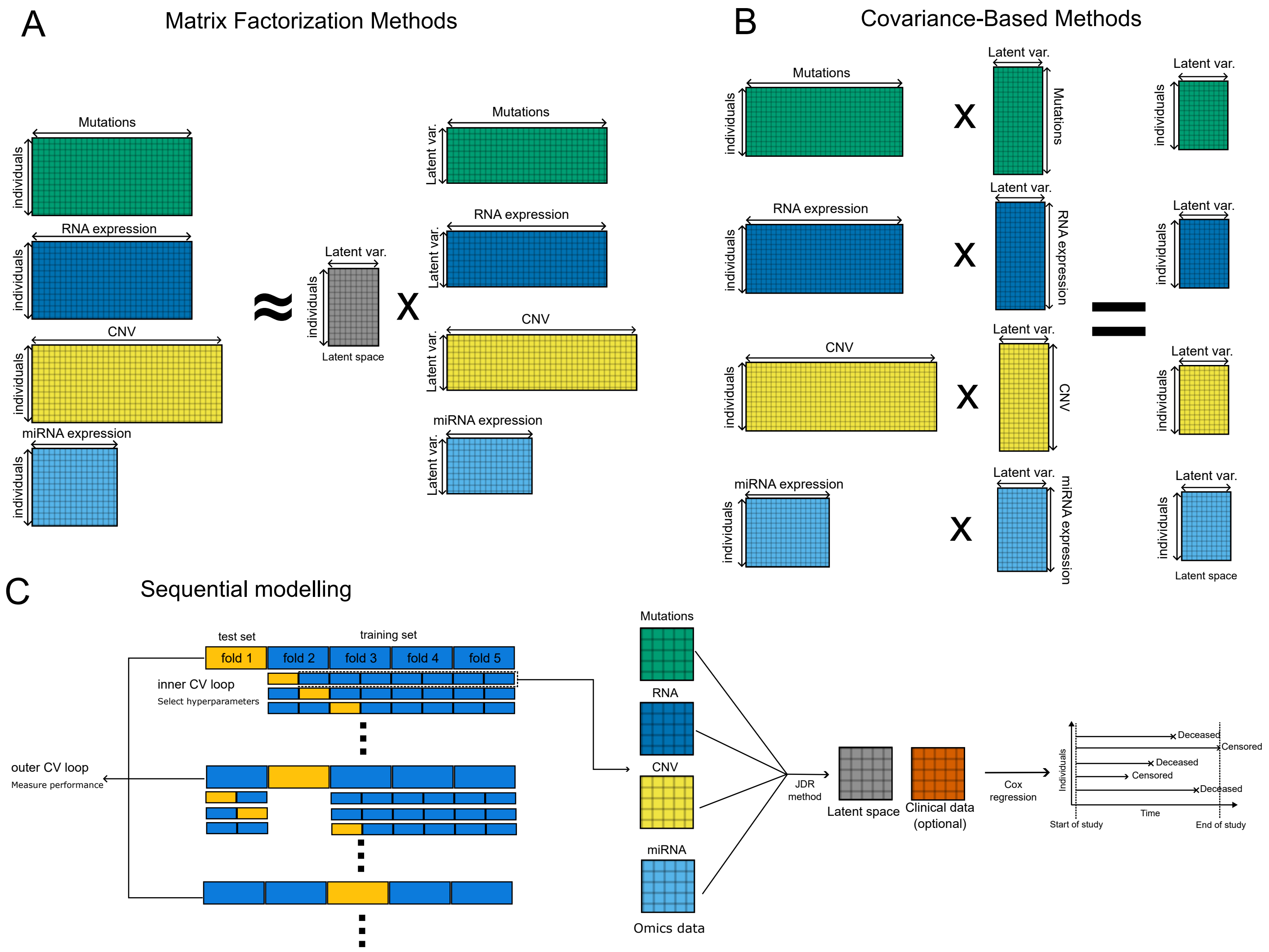

Figure 1: **Overview of JDR methods and cross-validation** we used for benchmarking JDR methods. Simplified representations of: A - Matrix Factorization methods, B - Covariance-Based methods, and C - sequential modeling for survival data and its cross-validation.

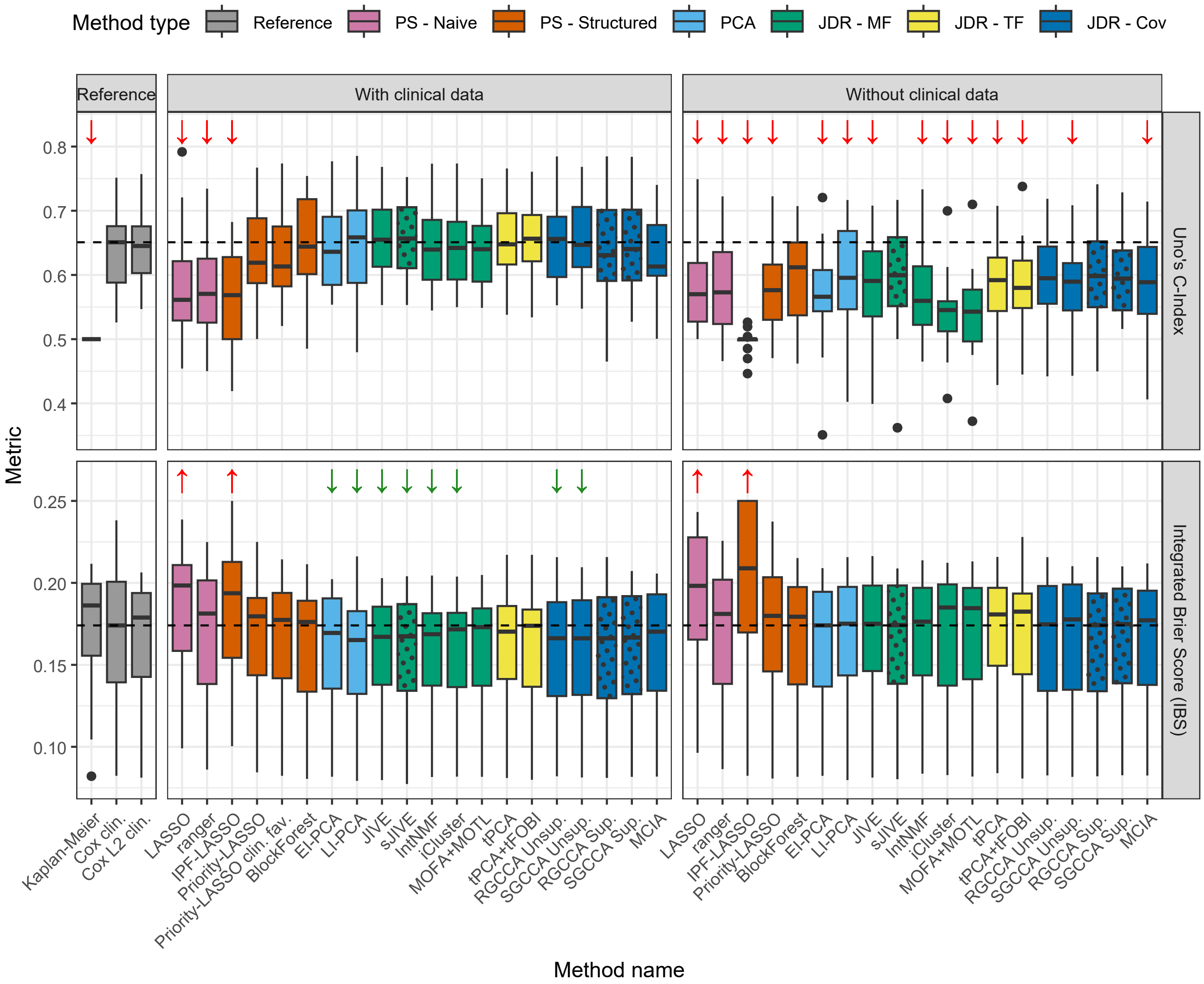

Figure 2: **Performances of each tested method across cancers**. For each cancer, the 50 outer folds were aggregated by taking the median. The top row corresponds to the C-Index (the higher the better) while the bottom row corresponds to the IBS (the lower the better). Columns depicts in order: reference methods that use either no variables (Kaplan-Meier estimator) or only clinical variables (Cox and regularized Cox models), methods with clinical data and methods without clinical data. Supervised JDR methods are dotted. Paired Wilcoxon tests were used to compare each methods to the baseline of the non-regularized Cox model using only clinical data. Pairing was done by cancer. P-values were corrected with the Benjamini-Hochberg procedure. An arrow in the direction of the effect is depicted above the corresponding boxplot when the corrected p-value is inferior to 0.05. A green arrow corresponds to an improvement, a red arrow shows a decrease in performance. PS: Previous Study [7].

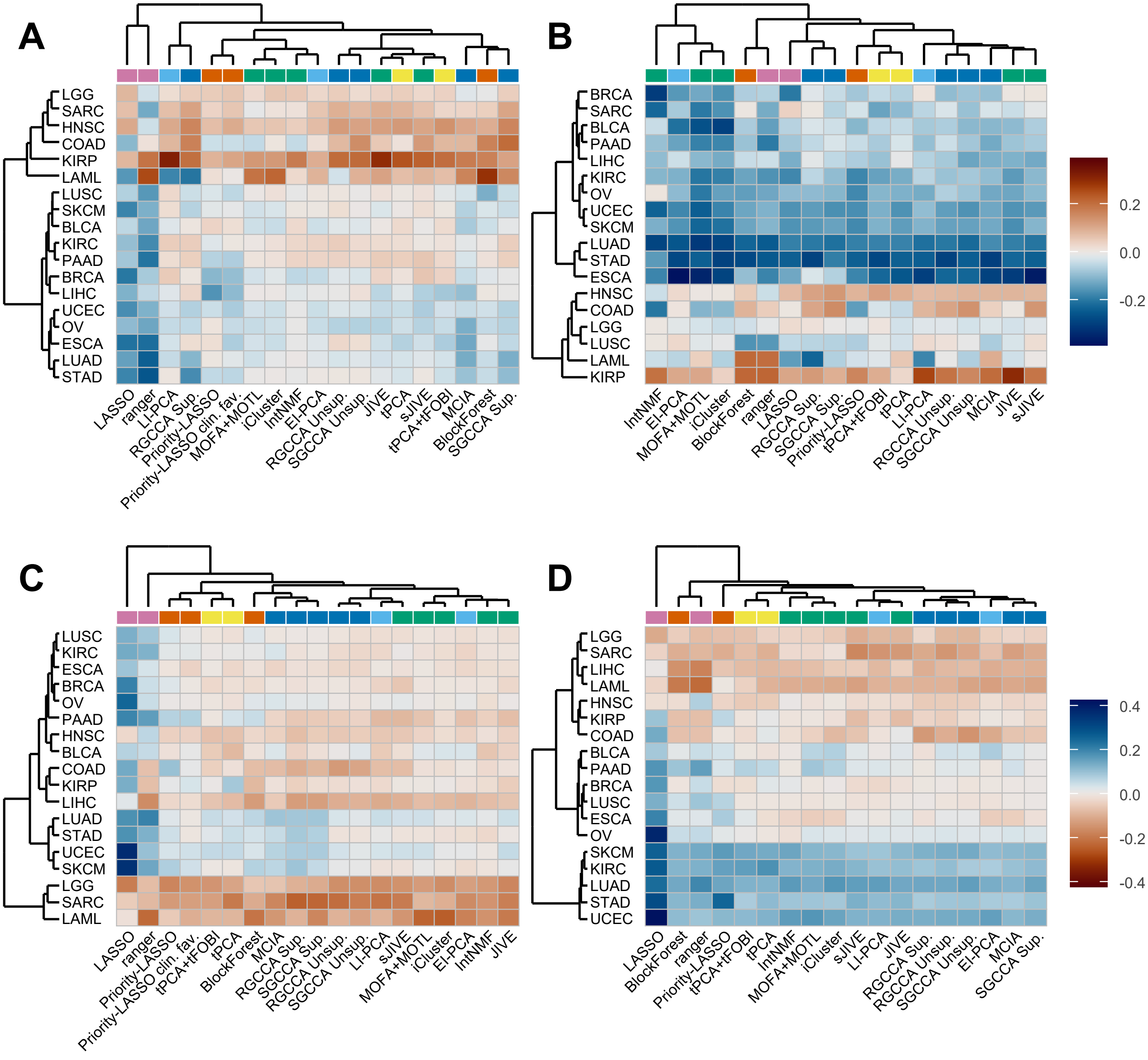

Figure 3: **Heatmaps of relative median performance** between each method and Cox clin. for each cancer. A - Uno's C-Index with clinical data. B - Uno's C-Index without clinical data. C - IBS with clinical data. D - IBS without clinical data. Red corresponds to an improvement while blue corresponds to a decrease in performance. Colors above the heatmaps correspond to method types as per figure 2. Clustering was done with ward.D2 linkage and euclidean distance. IPF-LASSO was removed from this analysis for two reasons: (i) it constantly underperforms enough to make other effects harder to see, which is explained by (ii) its high error rate, leading to missing survival performance values which are imputed to the default values of 0.5 for C-Index and 0.25 for IBS (see Methods 4.4 for more details).

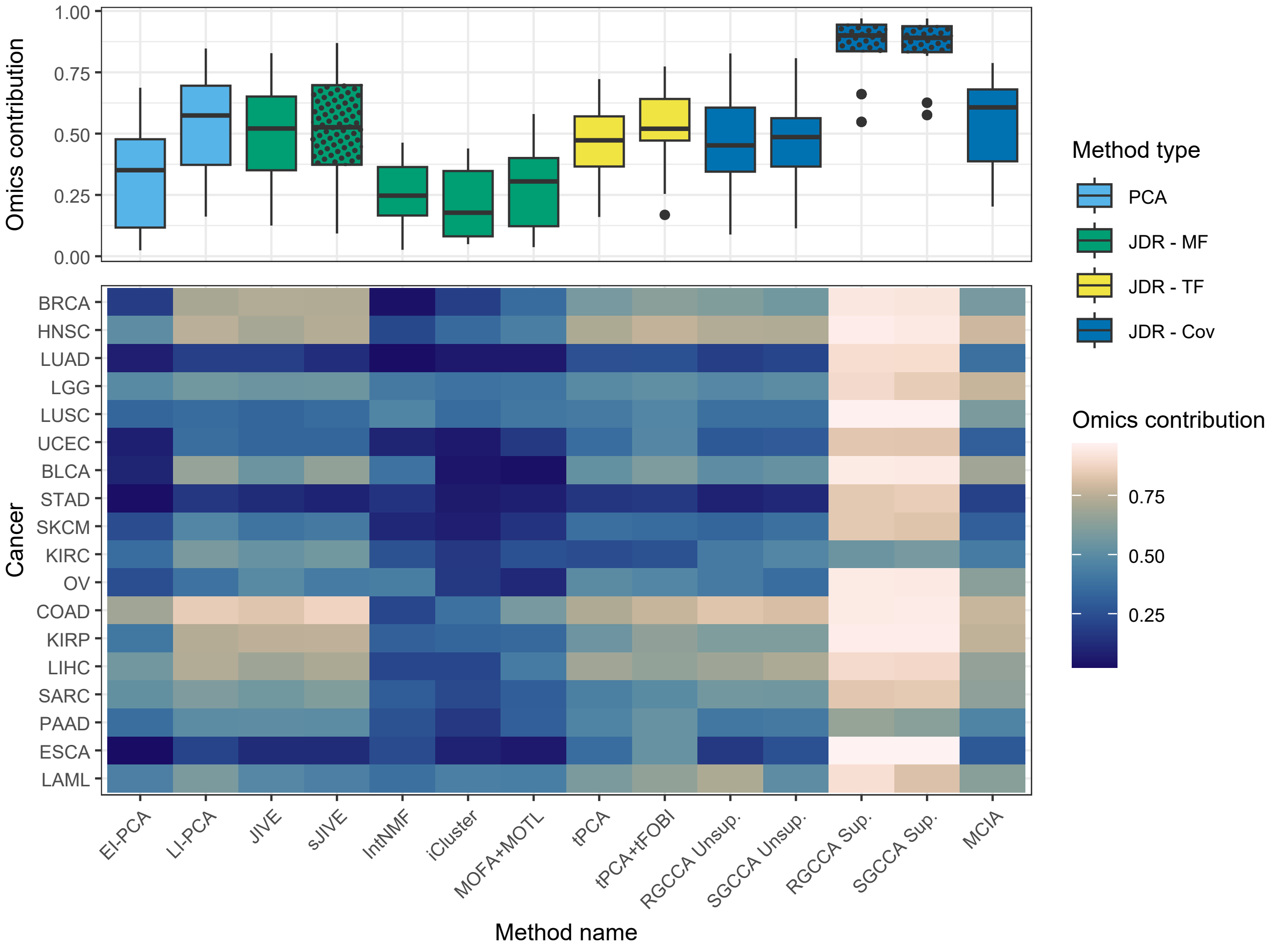

Figure 4: **Contribution of omics data to the prediction** of the Cox model, when clinical data are present. Above: Boxplot of mean contributions averaged over the 50 outer folds for each cancer. Below: heatmap of mean contribution per cancer and method. Supervised method are dotted.

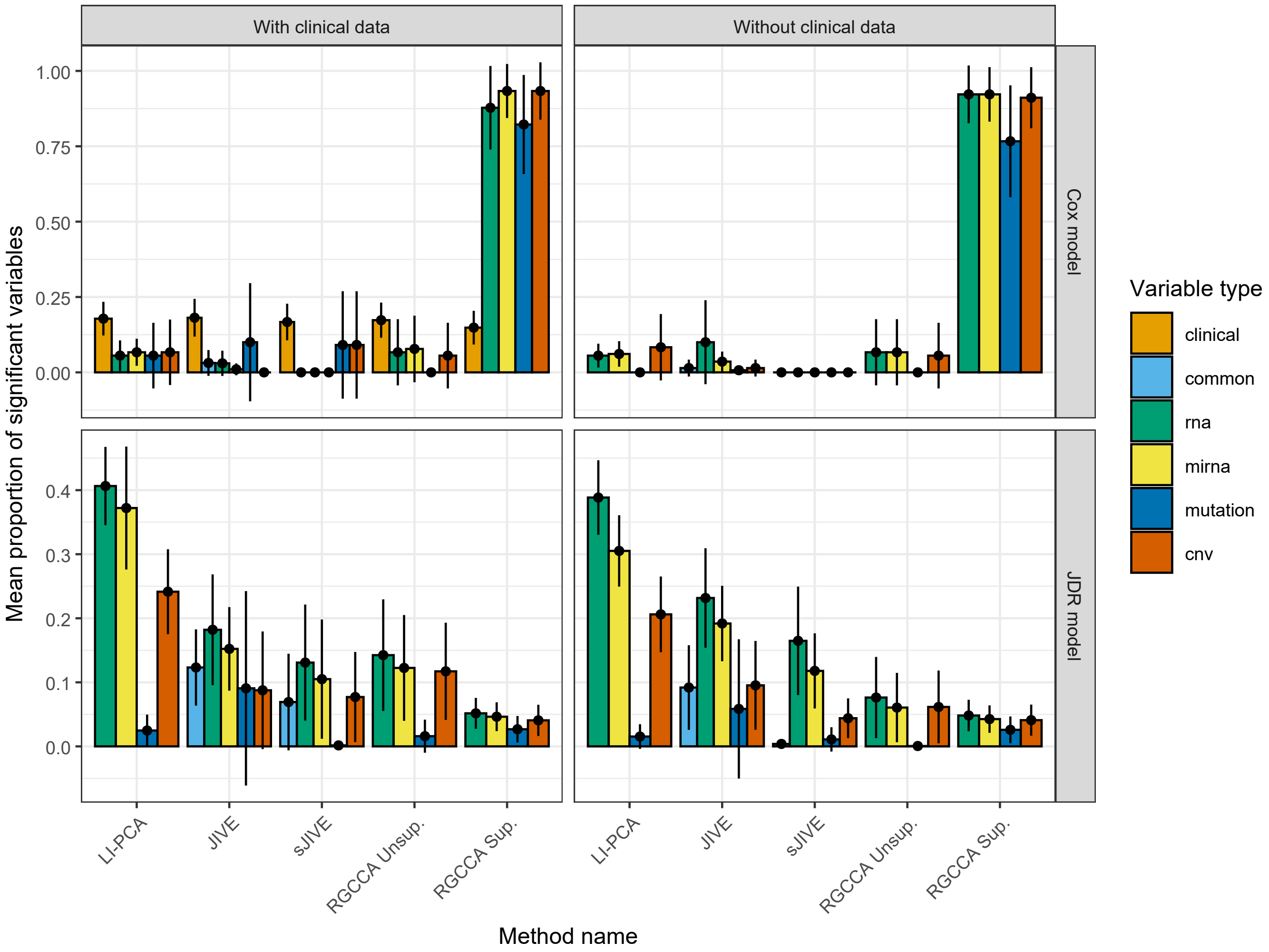

Figure 5: **Mean proportion of significant variables** across all 18 cancer for bootstrapped methods. The top and bottom rows depict respectively the proportion of significant variables for the Cox models and the JDR models, while the left and right columns depict respectively the models with and without clinical data added to the latent space. Errors bars depict the 95 % normal confidence interval of the mean for 18 cancers.

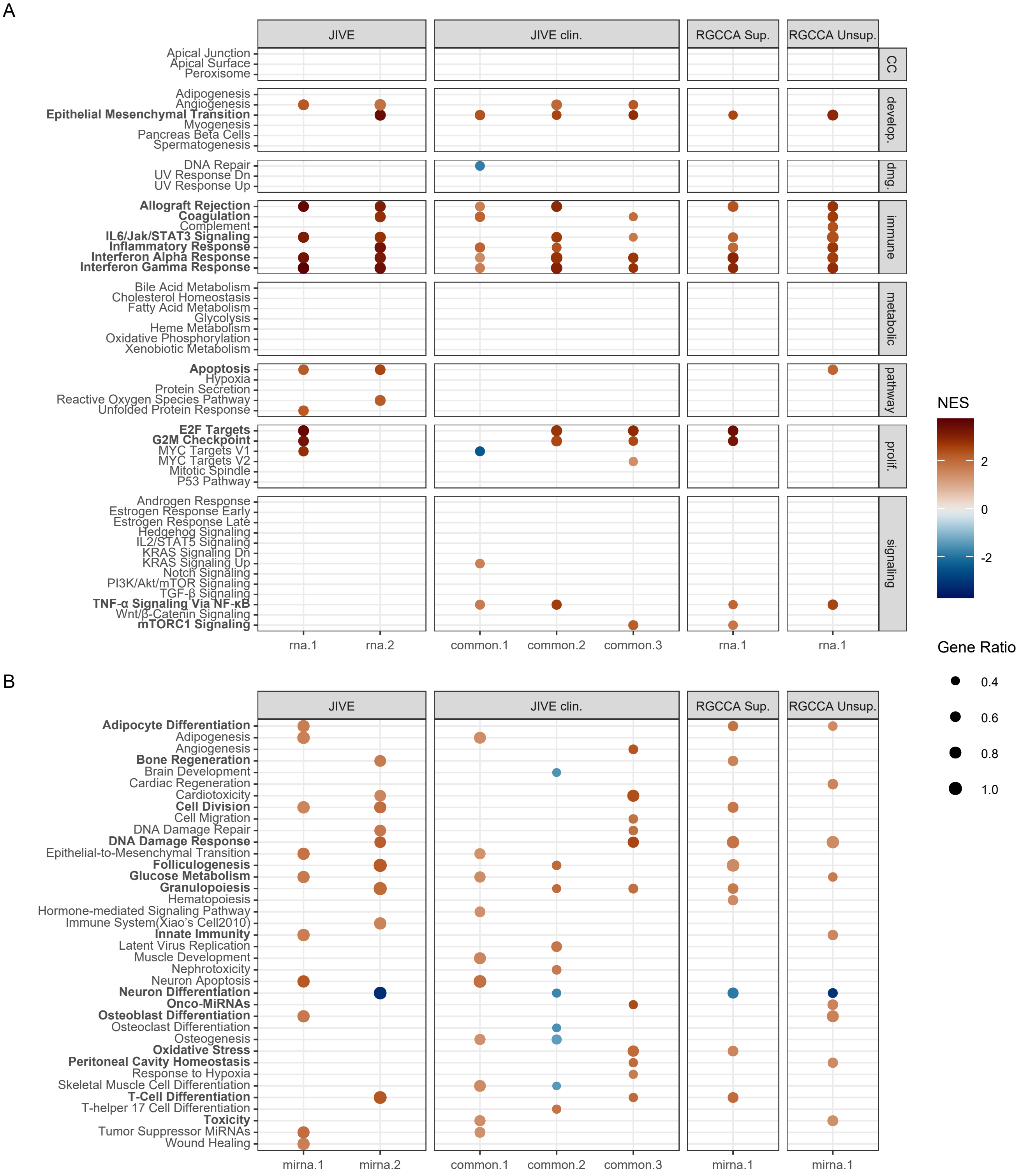

Figure 6: **Top 10 enriched gene sets** per component for JIVE, JIVE clin., RGCCA Sup. and RGCCA Unsup. on LGG, ranked by gene ratio. A - MSigDB hallmarks enrichment for RNA components. B - TAM 2.0 function gene sets enrichment for miRNA components. Only significantly enriched pathways were kept. block.n signifies that this represents the nth component for this block. NES signs were adjusted using the Cox's weight of their associated components such that a positive score corresponds to a deleterious effect on survival. Enrichment was done only for components significant for the Cox model. Gene sets in bold are retrieved by both at least one RGCCA and JIVE model. Hallmark categories are from Liberzon et al. [44]. NES: Normalized Enrichment Score, CC: Cellular Components, develop.: development, dmg: DNA damage, prolif.: proliferation.

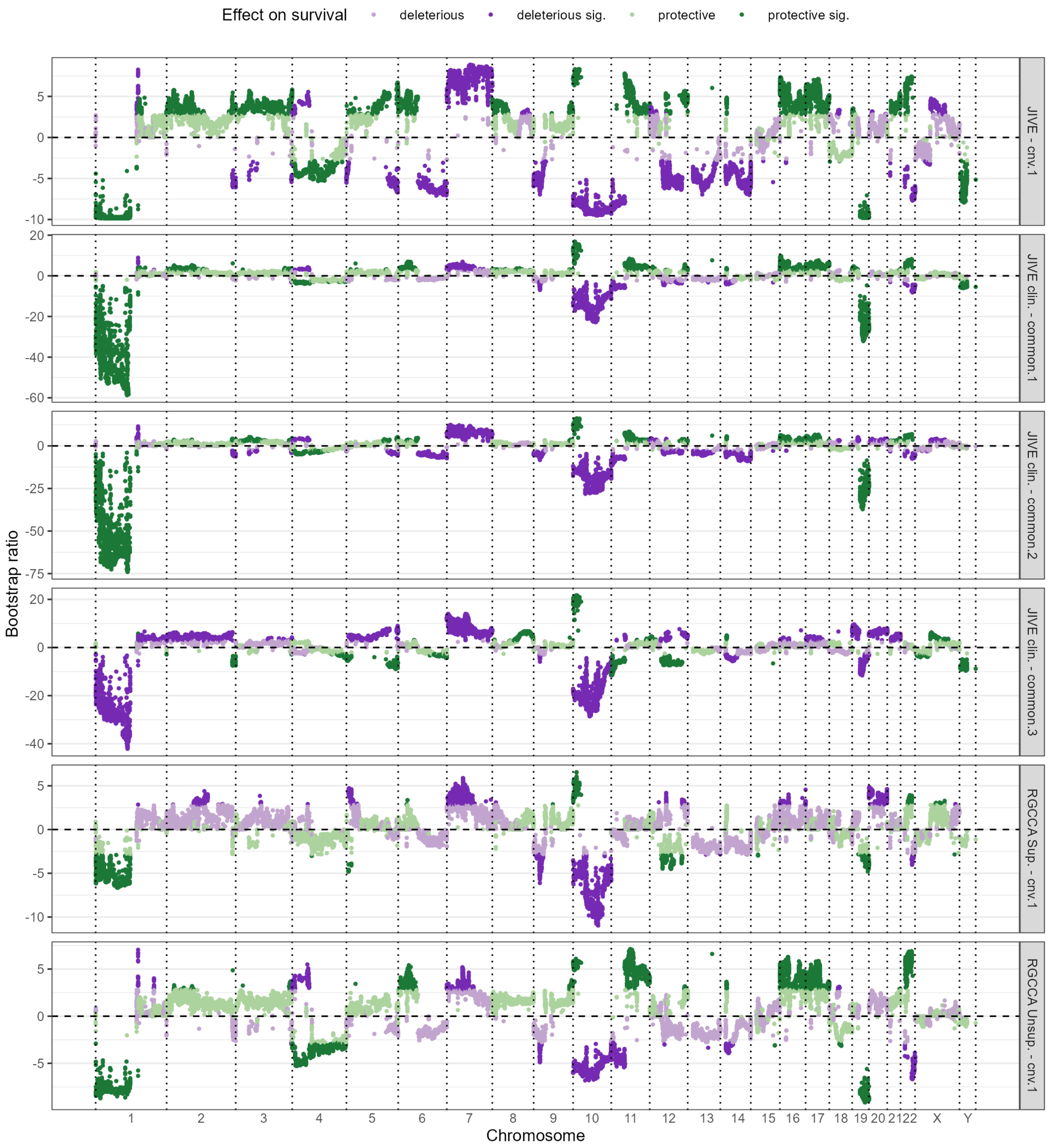

Figure 7: **Boostrap ratio of CNV components** for JIVE, JIVE clin., RGCCA Sup. and RGCCA Unsup. For each CNV variable, the sign of its represented bootstrap ratio is set to the sign of its mean value. The effect on survival is then determined by the product of the sign of its mean CNV value in the LGG dataset (main event deletion or amplification) of its bootstrap ratio (contribution of this variable to the whole component) and of the weight associated to the component by the Cox model (contribution of the whole component to survival). In the end, a positive product (resp. negative) corresponds to a deleterious (resp. protective) effect. Significance is determined with a Q-value threshold of 0.05.

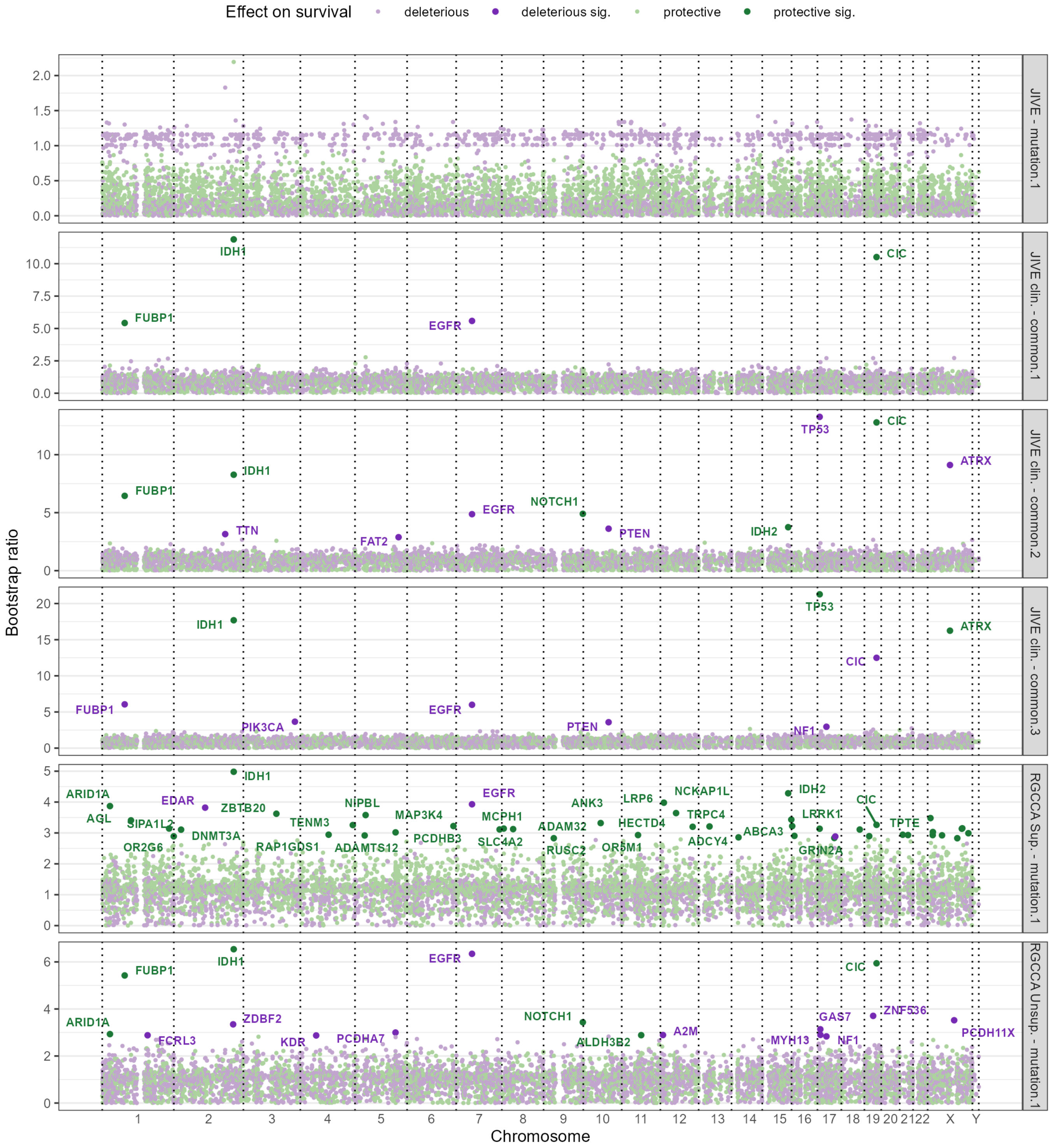

Figure 8: **Bootstrap ratio of mutation components** for JIVE, JIVE clin., RGCCA Sup. and RGCCA Unsup. For each gene, the absolute value of the bootstrap ratio is represented. The effect on survival is then determined by the product of the sign of its bootstrap ratio (contribution of this variable to the whole component) and of the weight associated to the component by the Cox model (contribution of the whole component to survival). In the end, a positive product (resp. negative) corresponds to a deleterious (resp. protective) effect. Significance is determined with a Q-value threshold of 0.05.